\documentclass[twocolumn,aps,prl,superscriptaddress,tightenlines]{revtex4}
\usepackage{amsmath}
\usepackage{amsfonts}
\usepackage{graphicx}
\usepackage{epsfig}
\usepackage{color}
\usepackage{textcomp}
\usepackage[colorlinks,citecolor=blue]{hyperref}

\begin{document}

\title{Cavity QED based on strongly localized modes: exponentially enhancing single-atom cooperativity}
\author{Qian Bin}
\affiliation{College of Physics, Sichuan University, Chengdu 610065, China}
\affiliation{School of Physics and Institute for Quantum Science and Engineering, Huazhong University of Science and Technology, and Wuhan Institute of Quantum Technology, Wuhan 430074, China}

\author{Ying Wu}
\affiliation{School of Physics and Institute for Quantum Science and Engineering, Huazhong University of Science and Technology, and Wuhan Institute of Quantum Technology, Wuhan 430074, China}

\author{Jin-Hua Gao}
\affiliation{School of Physics and Institute for Quantum Science and Engineering, Huazhong University of Science and Technology, and Wuhan Institute of Quantum Technology, Wuhan 430074, China}

\author{Aixi Chen}
\affiliation{Key Laboratory of Quantum State and Optical Field Manipulation of Zhejiang Province, Department of Physics, Zhejiang Sci-Tech University, Hangzhou 310018, China} 

\author{Franco Nori}
\affiliation{Theoretical Quantum Physics Laboratory, Cluster for Pioneering Research, RIKEN, Wako-shi, Saitama 351-0198, Japan}
\affiliation{Quantum Information Physics Theory Research Team, Quantum Computing Center, RIKEN, Wakoshi, Saitama, 351-0198, Japan}
\affiliation{Physics Department, The University of Michigan, Ann Arbor, Michigan 48109-1040, USA}

\author{Xin-You L\"{u}}
\email{xinyoulu@hust.edu.cn}
\affiliation{School of Physics and Institute for Quantum Science and Engineering, Huazhong University of Science and Technology, and Wuhan Institute of Quantum Technology, Wuhan 430074, China}
\date{\today}

\begin{abstract}
Large single-atom cooperativity in quantum systems is important for quantum information processing. Here, we propose to exponentially enhance the single-atom cooperativity parameter by exploiting the strongly localized effect of modes in cavity quantum electrodynamics  (QED) systems. By increasing the wing width of a cavity with special geometry symmetry, the interference property allows us to exponentially improves the quality factor $Q$ without altering the mode volume $V$  for cavities supporting subwavelength light modes. This effectively overcomes the trade-off between $Q$ and $V$ in conventional subwavelength Fabry–P\'{e}rot cavities. Consequently, we demonstrate the occurrence of ultra-long vacuum Rabi oscillations and the generation of strong photon blockade by enhancing the single-atom cooperativity parameter. This work offers a promising approach for advancing coherent manipulation and holds significant potential for applications in establishing longer-distance quantum communication networks, enhancing the precision and stability of quantum sensors, and improving the efficiency of quantum algorithms.
 
\end{abstract}
\maketitle

Strong coherent light-matter interactions are pivotal for applications ranging from quantum information processing\,\cite{Kimble2008} to precise sensing\,\cite{Degen2017RC}. These interactions, essential for enhancing the efficiency of coherent manipulation, have been a focus in cavity quantum electrodynamics (QED)\,\cite{Bose2014CC,Sipahigil2016ES, Evans2018BS, Liu2014LL,Lu2015WJ,Leroux2018GC,Qin2018ML, Sapienza2010TS, Qin2024KM, Huang2020Chen}. The single-atom cooperativity parameter (defined as $C=g^2/(\kappa\gamma)$, with $g$ representing the light-matter coupling strength, $\kappa$ the decay rate of cavity field, and $\gamma$ the atomic spontaneous emission rate) describes the balance between coherent interactions and dissipative processes within the cavity QED system, dictating the feasibility and efficiency of quantum manipulations. The value of $Q/V$, which relies on high finesse, is an important index for evaluating a cavity's performance in enhancing the single-atom cooperativity parameter. A high $Q/V$ ratio is desirable in many applications, such as lasers\,\cite{He2012OY,Jiang2016ZW} and sensors\,\cite{Alivisatos2004}. Over the past decades, great efforts have been made to reduce the mode volume $V$ and improve the quality factor $Q$ of cavities to enhance single-atom cooperativity in various systems, such as  microscopic Fabry-P\'{e}rot cavities\,\cite{Hunger2010SC, Greuter2014SN}, whispering-gallery-cavities\,\cite{Armani2003KS, Kippenberg2004SV}, hybrid photonic-plasmonic cavities\,\cite{Xiao2012LL, Min2009OS, Yin2016LB, Cognee2019DL, Conteduca2017RS, Xiong2022Xiao, Zhang2022LW}, and photonic crystal cavities\,\cite{Badolato20005HA, Englund2007FF,  Bose2012SK}. In nanocavity and microcavity QED systems, large cooperativity parameters—on the order of $10^6$ or even higher—can be reached  due to advancements in technology that easily achieve ultrahigh quality factors $Q$ or ultrasmall mode volumes $V$\,\cite{Choi2017HE, Najer2019SS, Albrechtsen2022LC}. 

Rydberg atoms have recently gained significant attention in quantum information processing due to their unique properties\,\cite{Gallagher2005, Saffman2010WM}, including strong dipole moments, long lifetimes, and ease of manipulation. However, the large radius of highly excited Rydberg atoms—often comparable to or larger than the wavelength of visible light—poses challenges for coupling with nanocavities and microcavities. Macroscopic cavities, such as millimeter- or centimeter-scale cavities, are generally better suited to the size of Rydberg atoms\,\cite{Guerlin2010BE, Parigi2012BS, Zhang2013SW, Haroche2020BR}. Nonetheless, most macroscopic cavity QED systems struggle to achieve high cooperativity parameters, as simultaneously attaining both high $Q$ and small $V$ in conventional large-scale cavities is difficult\,\cite{Svelto2010}.  Although it is possible to achieve high $Q$ and small $V$ through optimized designs such as photonic band gap structures\,\cite{Yun2001Chang, Armenise2010CC} and  inverse design techniques\,\cite{Molesky2018LP, Albrechtsen2022L}, they often come at the cost of increasing system complexity.   Therefore,  enhancing cooperative parameters in most large-scale cavity systems remains a challenge due to the  trade-off between high $Q$ and small $V$, but it is highly desirable for establishing long-time coherent manipulation.

Here, we present an approach to achieve a large single-atom cooperativity parameter and prolong light-matter coherent interaction within a specially designed Fabry–P\'{e}rot cavity, which is not constrained by the $Q$-$V$ trade-off.  By engineering the geometry of the cavity,  it becomes possible to significantly localize the subwavelength light modes while minimizing loss, thereby providing a platform for achieving large single-atom cooperativity. Specifically, by adding two wings to the Fabry–P\'{e}rot cavity and increasing the wing width, we can achieve an {\it exponentially enhancement of the quality factor $Q$ for strongly localized subwavelength modes without changing the mode volume $V$}. This overcomes the  trade-off between high $Q$ and small $V$ in conventional subwavelength Fabry–P\'{e}rot cavities,  leading to an exponential improvement in the $Q/V$ ratio and single-atom cooperativity parameter $C$  in cavity QED systems. This exponential improvement in $C$ can be achieved not only in idealized cavities but also in realistic structures with experimentally realizable designs.  Notably, a large single-atom cooperativity parameter $C$ can still be obtained even in the presence of mirror dissipation. Our results demonstrate that this cavity QED systems enables {\it ultra-long} vacuum Rabi oscillations and a {\it strong photon blockade} effect.  The proposed structure provides a new platform to overcome decoherence and prolong coherent manipulation of photons, offering significant potential application in quantum information processing.

\begin{figure}
\includegraphics[width=8.7cm]{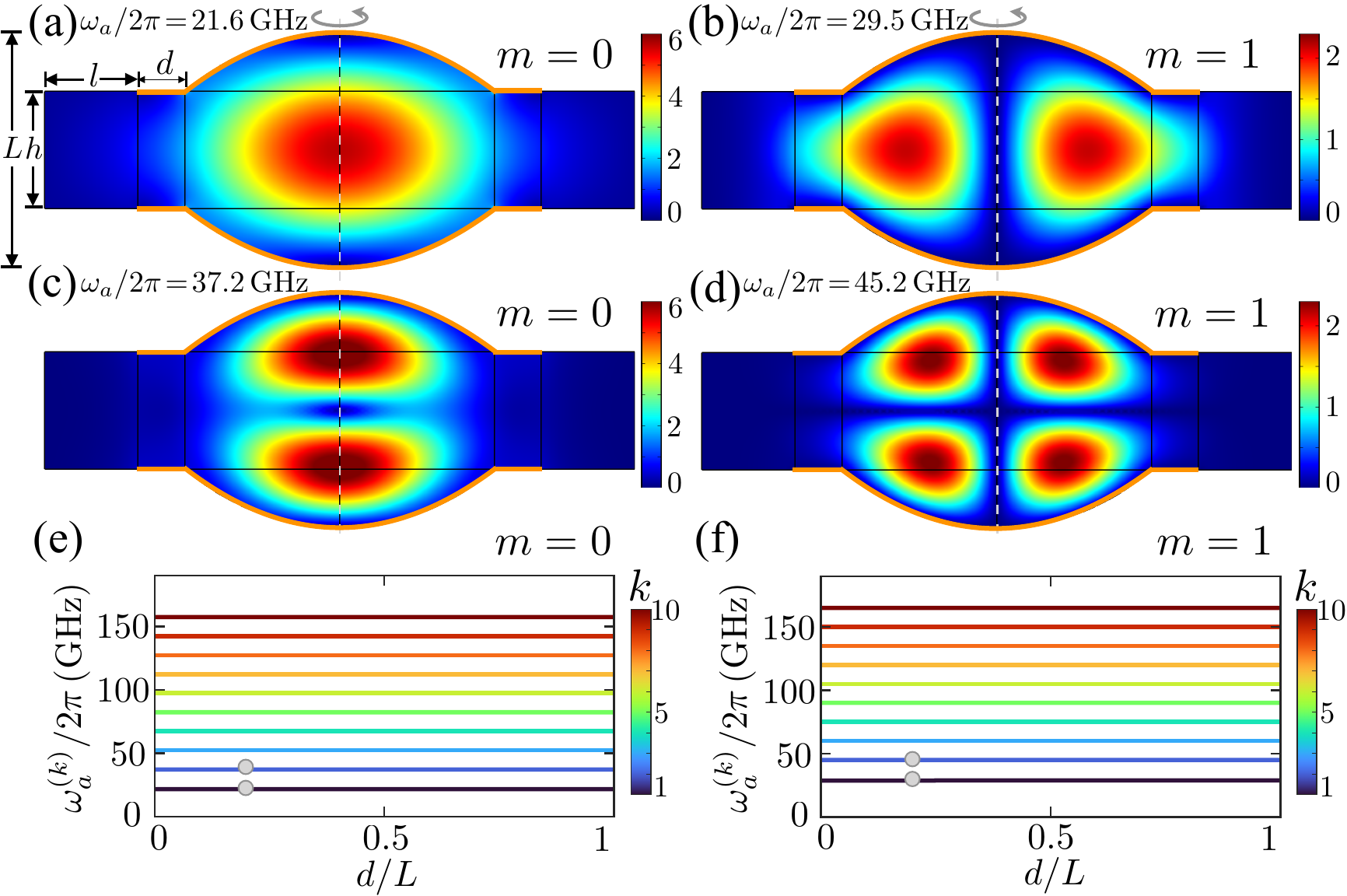}\\
\caption{(a-d) Electric field intensity distributions of different resonant modes for a specially designed Fabry–P\'{e}rot cavity with wing width $d=0.2L$, with corresponding resonant frequencies indicted by the gray dots in (e,f).  (e,f) Resonant frequencies $\omega_a^{(k)}/2\pi$ of the cavity versus $d/L$ for different  $m$ and  $k$. The geometric parameters of the cavities used here are: $L=1\,{\rm cm}$, $h=0.5 L$, and $l=0.4L$.}\label{fig1}
\end{figure}

\emph{Mode localization in the specially designed Fabry–P\'{e}rot cavity.}---As depicted in Fig.\,\ref{fig1}, we consider a centimeter-scale Fabry–P\'{e}rot cavity consisting of two reflected mirrors with two additional wings,  where the mirrors are shown by the yellow curves. The maximum and minimum distances between the reflected mirrors are $L$ and $h$, respectively, and $d$ represents the width of the wings and $l$ is the width of perfectly matched layers. The wing width $d= 0$ corresponds to a standard symmetric confocal cavity. To investigate the properties of the cavity, we preform finite-element simulations,  carried out under the condition of perfect electrical conductor (PEC) and utilize a  2D axisymmetric formulation. Figure \ref{fig1} shows the electric field intensity distributions of the modes with small longitudinal mode orders, which are primarily confined to the central region of the cavities rather than the wings.  Distributions of more modes are shown in the Supplemental Material\,\cite{supp}. Increasing the wing width $d$ does not significantly alter the eigenvalues of these modes, even for the longitudinal mode orders $k=1,2$, as evidenced by the eigenfrequency spectra in Figs.\,\ref{fig1}(e) and \ref{fig1}(f). The corresponding eigenfrequencies are $\omega_a^{(k)}\approx c[k+(m+1)/2]/(2\eta L)$, where $c$ is the speed of light in a vacuum, $\eta$ is the vacuum refractive index, and $k$ ($m$) represents the longitudinal (transverse) mode order. 

\begin{figure}
\includegraphics[width=8.7cm]{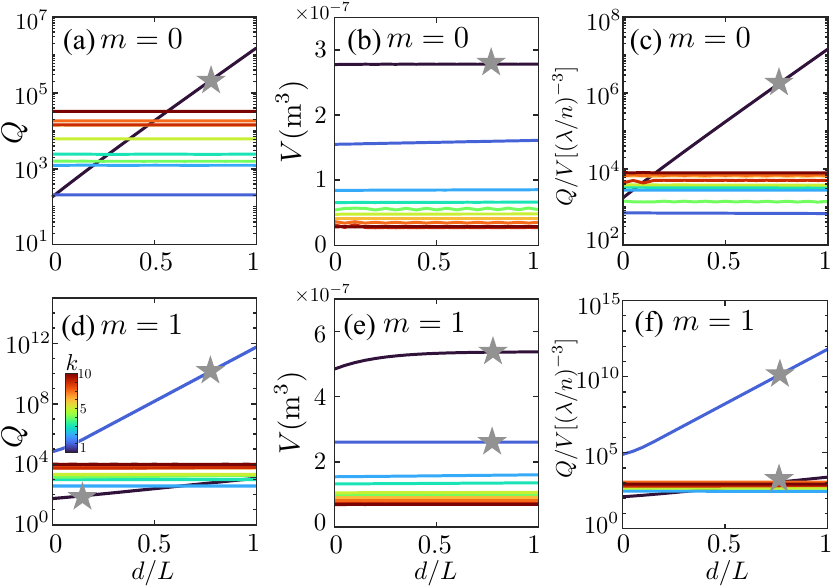}\\
\caption{(a,d) Quality factor $Q$, (b,e) mode volume $V$, and (c,f) the ratio $Q/V$ versus $d/L$ for the  Fabry–P\'{e}rot cavity in Fig.\,\ref{fig1} with different $m$ and $k$. Modes marked by asterisks indicate that the $Q$ increases drastically as $d/L$ increases, which correspond to the subwavelength light. Other system parameters are the same as in Fig.\,\ref{fig1}.}\label{fig2}
\end{figure}

Figure\,\ref{fig2} shows the effects of the wing width $d$ on the quality factor $Q$ and the mode volume $V$ of the cavity. Here $V$ can be calculated from  ${1}/{V}={\rm Re}\{{1}/{\nu_Q}\}$ with $\nu_Q={\langle \langle \tilde{{\bf f}}_c|\tilde{{\bf f}}_c\rangle \rangle}/[{\epsilon({\bf r}_c)  \tilde{{\bf f}}_c^2({\bf r}_c)}]$, where $\epsilon({\bf r})$ is the relative permittivity,  $ \tilde{{\bf f}}_c({\bf r}_c)$ is the eigenfunction of cavity mode, and ${\bf r}_c$ corresponds to the field maximum\,\cite{Kristensen2012VVH,  Kristensen2014Hughes, Kristensen2020HI}. For very large $Q$, the mode volume can be approximately reduced to $V=\int_{\mathcal{V}} { \epsilon({\bf r}) |{\bf E}({\bf r})|^2 }/{{\rm max}[\epsilon({\bf r}) |{\bf E}({\bf r})|^2 ]}d^3 {\bf r}$\,\cite{Spillane2005KV, Srinivasan2006BP}, where ${\bf E}({\bf r})$ represents the electric field, and $\mathcal{V}$ is the quantization volume of the electromagnetic field. The mode volume  $V$ remains unchanged with increasing wing width $d$, as the special geometry of the cavity enhances the electric-field intensities but hardly modifies the electric field distributions.  However, as indicated by the asterisks in Figs.\,\ref{fig2}(a) and \ref{fig2}(d), the quality-factor $Q$ improves almost exponentially with increasing $d$ for some subwavelenght light modes, where their wavelengths $\lambda\sim L$ and $\lambda^3\sim V$. Thus, an ultra-high $Q$ can be obtained in this subwavelength Fabry–P\'{e}rot cavity. From Figs.\,\ref{fig2}(a-b) and \ref{fig2}(d-e), it is also evident that the typical  trade-off between high $Q$ and small $V$, which exists in some macroscopic Fabry–P\'{e}rot cavity systems including conventional subwavelength Fabry–P\'{e}rot cavity\,\cite{Svelto2010} and plane-plane Fabry–P\'{e}rot cavity\,\cite{supp}, is overcome in this cavity.  This enables the realization of an ultrahigh figure of merit $Q/V$ in the cavity. As shown in Figs.\,\ref{fig2}(c) and \ref{fig2}(f),  the $Q/V$ ratio  improves almost exponentially with increasing $d$, and an ultrahigh $Q/V$  can be achieved by changing $d$. Achieving an ultra-high $Q/V$ is fundamentally important for obtaining large single-atom cooperativity in cavity QED systems. Here, we mainly utilize a winged symmetric confocal cavity  as an example to discuss the influence of adding two wings on the cavity properties. In the Supplemental Material\,\cite{supp}, we show that a mode in the winged microdome cavity still exhibits an almost exponential increase in quality factor $Q$ with increasing wing width, while its $Q$ value remains lower than that in the winged symmetric confocal Fabry–P\'{e}rot cavities due to weaker field confinement in the dome region and greater extension into the wings, leading to increased geometric losses.

 \begin{figure}
\includegraphics[width=8.7cm]{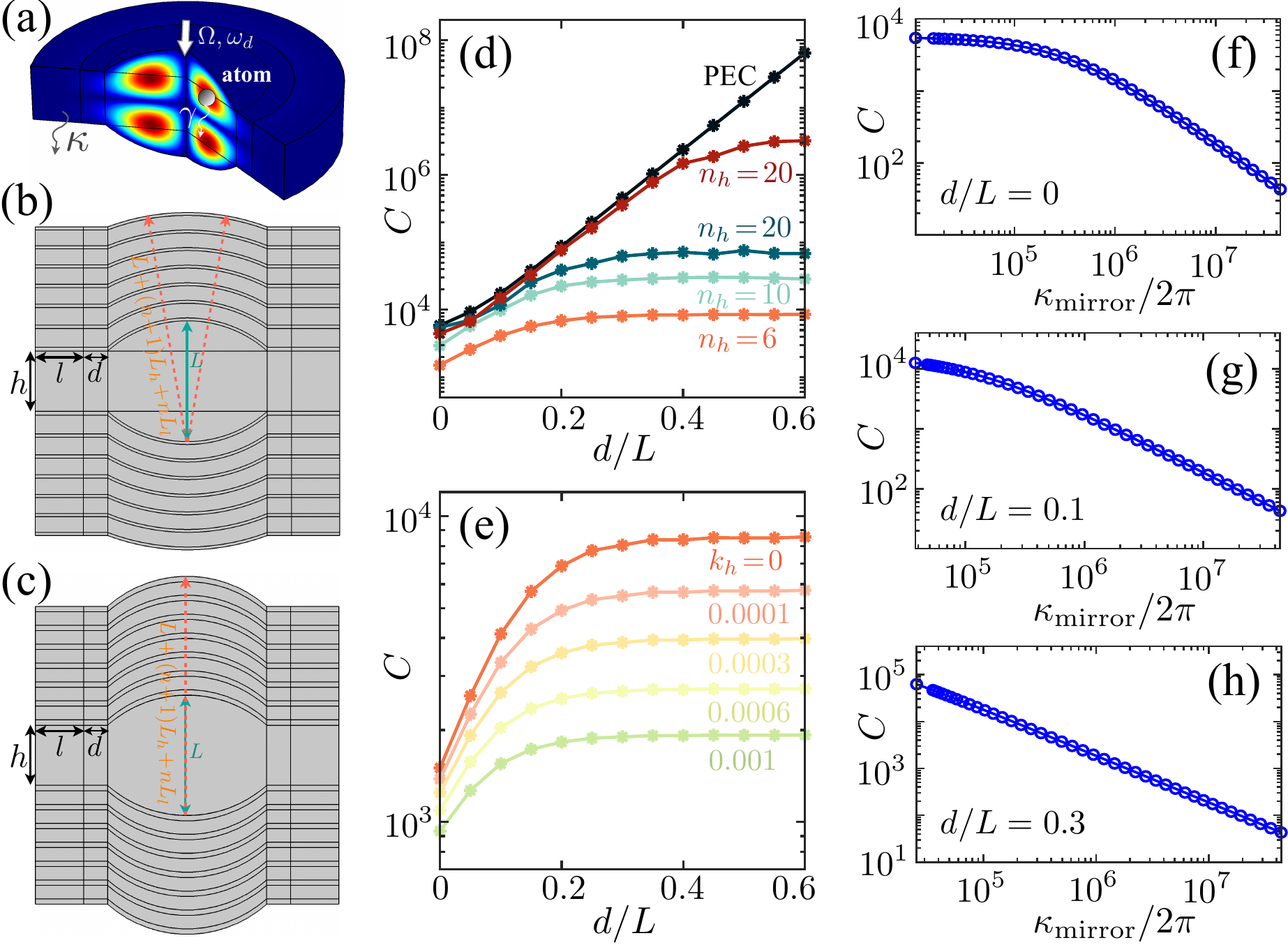}\\
\caption{(a) Schematic of the cavity QED system. (b,c) Special Fabry–P\'{e}rot cavities with ultrahigh reflectivity, where the mirrors are constructed by stacking layers of high- and low-refractive-index materials in different ways. (d,e) Single-atom cooperativity parameter $C$ versus  $d/L$ for different refractive indices $n_h$ and absorption coefficients $k_h$ of the high-index materials. (f-h) Dependence of $C$ on the mirror dissipation rate $\kappa_{\rm mirror}$ for different $d/L$. The black and red lines correspond to the cavity with a PEC and the cavity shown in (c), respectively; the remaining lines correspond to the cavity shown in (b). System parameters: (d) $n_l=1.2$, $k_h=k_l=0$, (e) $n_l=1.2$, $n_h=6$, $k_l=0.0001$, and other system parameters are the same as in Fig.\,\ref{fig1}.
} \label{fig3}
\end{figure}

\emph{Enhanced single-atom cooperativity in the cavity QED system}.---Now, we consider a Rydberg atom dipole coupled to the cavity to investigate the cavity QED system, as shown in Fig.\,\ref{fig3}(a).  The atom has a transition frequency $\omega_{\sigma}/2\pi=45.2\,{\rm GHz}$ and a transverse atomic dipole transition rate $\gamma/2\pi=2.5\times 10^3\,{\rm Hz}$. It is precisely placed at the location of maximum electric field strength. The coupling rate $g$ between the atom and a high-$Q$  cavity mode can be approximately expressed as $g=\gamma  \sqrt{ \mathbb{V}_a/V}$\,\cite{Vernooy1997Kimble, Buck2003Kimble}, where  $\mathbb{V}_a=3\pi c^3/(\gamma \omega_{\sigma}^2)$ is the characteristic atomic interaction volume. A more rigorous estimate of the local coupling strength and Purcell enhancement based on the Green's dyadic  is provided in the Supplemental Material, where the results are almost consistent with those from the approximation expression above\,\cite{supp}. Because $g\propto1/\sqrt{V}$, the coupling rate $g$ remains almost unchanged with increasing wing width $d$. However,  the single-atom cooperativity parameter $C=g^2/(\kappa\gamma)$ can be increased almost exponentially with $d$, since $\kappa\propto 1/Q$. This behavior is illustrated  by the black line in  Fig.\,\ref{fig3}(d), where the cavity mirrors are modeled as  PEC that fully reflect  light field. 

In realistic cavities, however, mirrors are never perfectly reflective.  We therefore consider the winged Fabry–P\'{e}rot cavities with experimentally realizable structures, in which the mirrors are constructed by alternately stacking dielectric materials with high and low refractive indices (RI), $n_h$ and $n_l$, respectively, as shown in Figs.\,\ref{fig3}(b,c).  The thickness $t_{h,l}$ of the high- and low-RI layers are determined by $n_h t_h=n_l t_l=\lambda/4$. These multilayer dielectric mirrors can achieve ultrahigh reflectivity, with the reflectivity improving as the contrast between $n_h$ and $n_l$,  and as more layers are added\,\cite{Stanley1994HO,Xu2019WC}.  Figures\,\ref{fig3}(d,e) show how the cooperativity parameter $C$ with  $d/L$ in such cavities.  Similar to the ideal case, increasing  $d$ leads to an exponential enhancement in $C$, but up to a plateau. Here, for simplicity, we focus exclusively on the cavity mode with $m=1$ and  $k=2$  (corresponding to the subwavelength light). This plateau arises primarily from imperfect mirror reflectivity. As the difference between $n_h$ and $n_l$ increases, and the absorption coefficients $k_h$ and $k_l$ of the high- and low-RI materials decrease, the onset of the plateau shifts to larger $d$ values, originating from the improved mirror performance. In the idealized limit where $n_h-n_l$ becomes very large and absorption loss is negligible, the plateau disappears entirely, and the trend of $C$ changing with $d$  approaches that of the PEC-mirror case. However, high-RI materials in the microwave regime often suffer from significant dielectric losses due to intrinsic absorption.  We can thus consider materials such as ${\rm LaAlO_3}$, ${\rm BaTiO_3}$,  ${\rm CaTiO_3}$, or Fe-doped ceramics for the high refractive indices layers, and Polystyrene Foam, PTFE, Rohacell, Airex for the low refractive indices layers\,\cite{Trichet2016TD,Gaponenkoa2019KR, Shihab2020AM}. For this cavity structure, the cooperativity parameter $C$  can reach value around  $4000$ when $d=0.4L$, even for parameters such as $n_h=6, n_l=1.2$ and $k_h=0.0003, k_l=0.0001$. In Figs.\,\ref{fig3}(f-h), we show the effect of the dissipation rate of mirrors $\kappa_{\rm mirror}$ on $C$, which demonstrate that even with a large dissipation rate $\kappa_{\rm mirror}/2\pi=1\,{\rm MHz}$, a large single-atom cooperativity parameter $C$ exceeding $10^{3}$ can be achieved in our cavity QED system when $d=0.3L$.


%
\begin{figure}
\includegraphics[width=8.7cm]{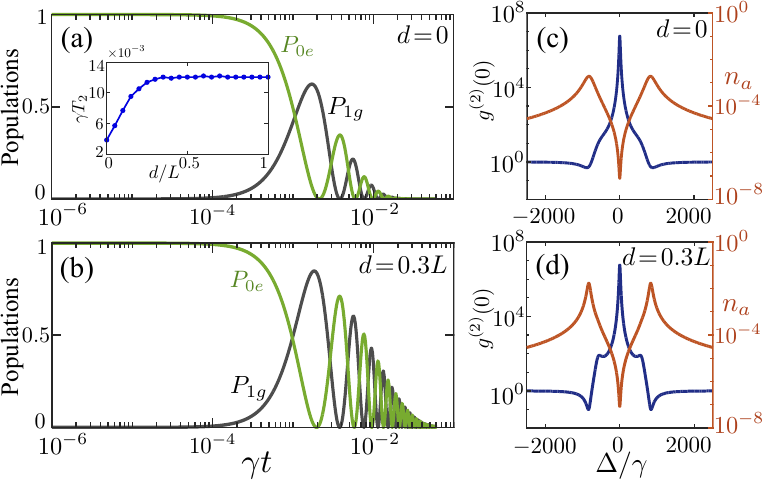}\\
\caption{(a,b) Population dynamics of the system states, $P_{0e}=|\langle 0, e|\psi(t)\rangle|^2$ and  $P_{1g}=|\langle 1, g|\psi(t)\rangle|^2$, for different  $d/L$ in the absence of driving. Inset: The decoherence time $T_2$ versus $d/L$, where $T_2$ represents the time for the amplitude to decay to $1/e$.  
(c,d) Zero-time delay second-order photon correlations $g^{(2)}(0)$ (blue lines) and mean photon numbers $n_a$ (red lines) versus  $\Delta/\gamma$ for different $d/L$ in the presence of driving. The cavity chosen here corresponds to Fig.\,\ref{fig3}(b) and the mode with $m=1$ and  $k=2$. System parameters used here are: $\omega_a/2\pi=\omega_{\sigma}/2\pi=45.2\,{\rm GHz}$, $\gamma/2\pi=2.5\times10^3~{\rm Hz}$, $\Omega=12\gamma$,  $n_h=6$, $n_l=1.2$,  $k_h=0.0003$, and $k_l=0.0001$.}\label{fig4}
\end{figure}

\emph{Ultra-long vacuum Rabi oscillations and strong photon blockade.}---Next, we demonstrate the realization of large single-atom cooperativity in a cavity QED system, where the cavity is constructed using the multilayer dielectric mirrors discussed above, and explore its applications through vacuum Rabi oscillations and single-photon blockade. The Hamiltonian of the cavity QED system under the rotating wave approximation  is 
\begin{align}\label{eq2}
 &H=\omega_a a^\dag a +\omega_{\sigma}\sigma^\dag \sigma+g(a^\dag\sigma+a\sigma^\dag), 
 \end{align}
 where $a$ ($a^\dag$) is the annihilation (creation) operator of the cavity mode with resonant frequency $\omega_a$, and $\sigma^\dag$($\sigma$) is the raising (lowering) operator of the atom in the two-level basis $\{|e\rangle, |g\rangle \}$, with $|e\rangle$ ($|g\rangle$) being the excited (ground) state of the atom. Taking system dissipations into account, the dynamics of the system can be described by the quantum master equation\,\cite{Gu2017KM, Kockum2019MD,FornDiaz2019LR}
\begin{align}\label{eq3}
 &\frac{d}{dt} \rho=-i[H,\rho]+\kappa \mathcal{L}[a]\rho+\gamma \mathcal{L}[\sigma]\rho, 
 \end{align}
where $\mathcal{L}[o]\rho=(2o\rho o^\dag-\rho o^\dag o-o^\dag o\rho)/2$ is the Lindblad superoperator, and $\rho$ is the systems's density operator. By numerically solving Eq.\,(\ref{eq3}), we can obtain the exact results of dynamical evolution of the dissipative system $P_{0e}(t)=|\langle 0, e|\psi(t)\rangle|^2$ and  $P_{1g}(t)=|\langle 1, g|\psi(t)\rangle|^2$, as shown in Figs.\,\ref{fig4}(a,b). Compared to Figs.\,\ref{fig4}(a) and \ref{fig4}(b),  it is clear that the period of the vacuum Rabi oscillations increases with increasing $d$.  The weaker the decay rate of the cavity, the longer the period of energy exchange becomes, with $d=0$ and $d=0.3L$ corresponding to the cavity decay rate $\kappa/2\pi=1.4\times 10^6~{\rm Hz}$ and $\kappa/2\pi=4.5\times 10^5~{\rm Hz}$, respectively. In the inset of Fig.\,\ref{fig4}(a), we also show the decoherence time $T_2$ of the dissipative cavity QED system, where $T_2$ represents the time for the amplitude to decay to $1/e$. The decoherence time $T_2$  improves almost exponentially with the wing width $d$, until it reaches a plateau, consistent with the behavior of the cooperativity shown in Fig.\,\ref{fig3}(e). These results demonstrate that strong, effective coherent light-matter coupling and large single-atom cooperativity in the cavity QED system can be significantly enhanced by changing the wing width of the winged Fabry–P\'{e}rot cavity.


The enhanced effective light-matter coupling and large single-atom cooperativity offers significant potential applications for single-photon\,\cite{Birnbaum2005BM} and few-photon manipulation\,\cite{Miranowicz2013PL, Hamsen2017TW}.  For instance,  the single-photon blockade effect occurs when the system exhibits effective strong nonlinearity in a cavity QED system\,\cite{Birnbaum2005BM}. Here, single-photon blockade means that the presence of a single photon in a cavity can block the excitation of other photons due to the strong effective nonlinearity, serving as one of the mechanisms to realize a single-photon source\,\cite{Lounis2005Orrit}. Single-photon blockade can be quantitatively characterized by the zero-time delay second-order photon correlation function $g^{(2)}(0)=\langle a^{\dag}  a^{\dag}  a a\rangle/\langle a^\dag a\rangle^2$\,\cite{Glauber1963}, where the condition $g^{(2)}(0)\to0$ indicates a relatively strong single-photon blockade. The smaller the correlation  $g^{(2)}(0)$, the stronger the single-photon blockade effect, implying a higher purity of the single photon in the cavity. The effective nonlinearity strength of the dissipative cavity QED system can be defined as $\mathcal{N}_{\rm eff}=\alpha/\beta$, with $\alpha=\sqrt{(\kappa-\gamma)^2-16lg^2}$ and $\beta=(2l-1)\kappa+\gamma$, which is derived from the effective eigenfrequencies of the $l$th-fold JC ladder $\omega_{\rm eff}^{n}=l\omega_a-i\beta/4\pm i\alpha/4$, by calculating the effective Hamiltonian of the dissipative system $H_{\rm eff}=(\omega_a-i\kappa/2) a^\dag a +(\omega_{\sigma}-i\gamma/2)\sigma^\dag \sigma+g(a^\dag\sigma+a\sigma^\dag)$\,\cite{Scully1997Zubairy, Agarwal2013}. 

 Considering the cavity is driven by a laser with frequency $\omega_d$ and amplitude $\Omega$, then the total system Hamiltonian is $H_t=H+\Omega (a^\dag e^{-i\omega_d t}+a e^{i\omega_d t} )$. In the frame rotating at the frequency $\omega_d$, the total Hamiltonian becomes $H_t=\Delta a^\dag a+\Delta\sigma^\dag \sigma+g (a^\dag \sigma+a\sigma^\dag)+\Omega(a^\dag +a)$,
where $\Delta=\omega_a-\omega_d=\omega_{\sigma}-\omega_d$ is the detunings of the cavity (and atom) frequency with respect to the laser driving. We calculate the exact numerical results of the photon statistics of the system in the steady-state ($t\to \infty$) through solving Eq.\,(\ref{eq3})\,\cite{Johansson2012NN}, with the results shown in Figs.\,\ref{fig4}(c,d). In Fig.\,\ref{fig4}(c) with $d=0$, the correlation function has a small value of $g^{(2)}(0)\approx0.5$,  but the corresponding mean photon number is also very low, with $n_a\approx 0.002$. Increasing the wing width to $d=0.3L$, as shown in Fig.\,\ref{fig4}(d), the photon blockade effect with  $g^{(2)}(0)\approx 0.1$ and $n_a\approx0.017$ occurs at a detuning  $\Delta/\gamma\approx\pm832$. This is because the increase of $d$ reduces the cavity decay rate $\kappa$ and enhances the effective nonlinearity $\mathcal{N}_{\rm eff}$ of the dissipative system.   These results demonstrate that increasing the wing width $d$ can simultaneously reduce the correlation $g^{(2)}(0)$ and improve the mean photon number $n_a$ of the steady state.  The enhanced purity and brightness of single photons in the system hold promise for the generation of an ideal single-photon source\,\cite{Lounis2005Orrit}.

\emph{Conclusions}.---In summary, we have proposed a method to exponentially enhance single-atom cooperativity in a cavity QED system. Due to the strongly localized effect of cavity modes with subwavelength light, increasing the width of the two wings of the cavity does not change the mode volume $V$ but exponentially improves the quality factor $Q$ of the cavity modes at small  longitudinal  mode orders with subwavelength light. This breakthrough overcomes the  trade-off between high $Q$ and small $V$, allowing for an exponentially enhancement of the single-atom cooperativity parameter with increasing the wing width. The cavity QED system, consisting of a  cavity coupled to an atom, enables the occurrence of ultra-long vacuum Rabi oscillation and strong single-photon blockade. Such advanced coherent manipulation could significantly enhance the computational power and accuracy of quantum computers\,\cite{Ladd2010JL} by stabilizing qubits and increasing the efficiency of quantum algorithms. Furthermore, it may aid in establishing longer-distance quantum communication networks by enhancing the security and reach of information transmission\,\cite{Gisin2002Thew, Bassoli2021BD}, improve the precision and stability of quantum sensors, and enable high-quality quantum imaging even under low-light conditions\,\cite{Kolobov2007, Magana-Loaiza2019Boyd}.  

\emph{Data availability}.---The data that support the findings of this Letter are not publicly available. The data are available from the authors upon reasonable request.

We thank Profs.\,T.\,Liu and S.\,Hughes for helpful discussions. This work is supported by the National Key Research and Development Program of China (Grant No.\,2021YFA1400700),  the National Science Fund for Distinguished Young Scholars of China (Grant No.\,12425502), the National Natural Science Foundation of China (Grant No.\,12205109), the Fundamental Research Funds for the Central Universities (Grant No.\,2024BRA001), and the Sichuan Science and Technology Program(Grant No.\,2025ZNSFSC0057). F. N. is supported in part by Nippon Telegraph and Telephone Corporation (NTT) Research, the Japan Science and Technology Agency (JST) [via the Quantum Leap Flagship Program (Q-LEAP), and the Moonshot R$\&$D Grant No. JPMJMS2061], the Asian Office of Aerospace Research and Development (AOARD) (via Grant No. FA2386-20-1-4069), and the Office of Naval Research (ONR) (via Grant No. N62909-23-1-2074).The computation is completed in the HPC Platform of Huazhong University of Science and Technology.


\begin{thebibliography}{4}
\bibitem{Kimble2008} H.J. Kimble, The quantum internet, Nature {\bf 453}, 1023 (2008).

\bibitem{Degen2017RC} C.L. Degen, F. Reinhard, and P. Cappellaro, Quantum sensing, Rev. Mod. Phys. {\bf 89}, 035002 (2017).

\bibitem{Sapienza2010TS} L. Sapienza, H. Thyrrestrup, S. Stobbe, P. D. Garcia, S. Smolka, P. Lodahl, Cavity Quantum Electrodynamics with Anderson-Localized Modes, Science {\bf 327}, 1352 (2010). 


\bibitem{Liu2014LL} Y.-C. Liu, X. Luan, H.-K. Li, Q. Gong, C.W. Wong, and Y.-F. Xiao, Coherent Polariton Dynamics in Coupled Highly Dissipative Cavities, Phys. Rev. Lett. {\bf 112}, 213602 (2014).

\bibitem{Bose2014CC}  R. Bose, T. Cai, K.R. Choudhury, G.S. Solomon, and E. Waks, All-optical coherent control of vacuum Rabi oscillations, Nat. Photonics {\bf 8}, 858 (2014)

\bibitem{Lu2015WJ} X.-Y. L\"{u}, Y. Wu, J.R. Johansson, H. Jing, J. Zhang, and F. Nori, Squeezed Optomechanics with Phase-Matched Amplification and Dissipation, Phys. Rev. Lett. {\bf 114}, 093602 (2015).

\bibitem{Sipahigil2016ES} A. Sipahigil, R. E. Evans, D. D. Sukachev, M. J. Burek, J. Borregaard, M. K. Bhaskar, C. T. Nguyen, J. L. Pacheco, H. A. Atikian, C. Meuwly, R. M. Camacho, F. Jelezko, E. Bielejec, H. Park, M. Loncar, and M. D. Lukin, Science {\bf 354}, 847 (2016).

\bibitem{Qin2018ML} W. Qin, A. Miranowicz, P.-B. Li, X.-Y. L\"{u}, J.Q. You, and F. Nori, Exponentially Enhanced Light-Matter Interaction, Cooperativities, and Steady-State Entanglement Using Parametric Amplification, Phys. Rev. Lett. {\bf 120}, 093601(2018).

\bibitem{Leroux2018GC} C. Leroux, L.C.G. Govia, and A.A. Clerk, Enhancing Cavity Quantum Electrodynamics via Antisqueezing: Synthetic Ultrastrong Coupling, Phys. Rev. Lett. {\bf 120}, 093602 (2018).

\bibitem{Evans2018BS}  R. E. Evans, M. K. Bhaskar, D. D. Sukachev, C. T. Nguyen, A. Sipahigil, M. J. Burek, B. Machielse, G. H. Zhang, A. S. Zibrov, E. Bielejec, H. Park, M. Loncar, and M. D. Lukin, Science {\bf 362}, 662 (2018).

\bibitem{Huang2020Chen} S. Huang and A. Chen, Mechanical squeezing in a dissipative optomechanical system with an optical parametric amplifier, Phys. Rev. A {\bf 102}, 023503 (2020).

\bibitem{Qin2024KM} W. Qin, A.F. Kockum, C.S. Mu\~{n}oz, et al., Quantum amplification and simulation of strong and ultrastrong coupling of light and matter, Phys. Rep. {\bf 1078}, 1-59 (2024).

\bibitem{He2012OY} L. He, \c{S}. K. \"Ozdemir, and L. Yang, Whispering gallery microcavity lasers, Laser  Photonics Rev. {\bf 7}, 60-82 (2012).

\bibitem{Jiang2016ZW} X.-F. Jiang, C.-L. Zou, L. Wang, Q.H. Gong, and Y.-F. Xiao, Whispering-gallery microcavities with unidirectional laser emission, Laser  Photonics Rev. {\bf 10}, 40-61 (2016).

\bibitem{Alivisatos2004} P. Alivisatos, The use of nanocrystals in biological detection, Nat. Biotechnol. {\bf 22}, 47 (2004).

\bibitem{Hunger2010SC} D. Hunger, T. Steinmetz, Y. Colombe, C. Deutsch, T. W. H\"{a}nsch, and J. Reichel, A fiber Fabry–Perot cavity with high finesse, New J. Phys. 12 065038 (2010).

\bibitem{Greuter2014SN}  L. Greuter, S. Starosielec, D. Najer, A. Ludwig, L. Duempelmann, D. Rohner, R.J. Warburton, A small mode volume tunable microcavity: Development and characterization, Appl. Phys. Lett. {\bf 105}, 121105 (2014). 




\bibitem{Armani2003KS} D. K. Armani, T. J. Kippenberg, S. M. Spillane, and K. J. Vahala, Ultra-high-Q toroid microcavity on a chip, Nature  {\bf 421}, 925 (2003).
 
\bibitem{Kippenberg2004SV} T. J. Kippenberg, S. M. Spillane,  K. J. Vahala,  Demonstration of ultra-high-Q small mode volume toroid microcavities on a chip, Appl. Phys. Lett. {\bf 85}, 6113 (2004).


\bibitem{Min2009OS} B. Min, E. Ostby, V. Sorger, E. Ulin-Avila, L. Yang, X. Zhang, and K. Vahala, High-Q surface-plasmon-polariton whispering-gallery microcavity, Nature {\bf 457}, 455 (2009).

\bibitem{Xiao2012LL} Y.-F. Xiao, Y.-C. Liu, B.-B. Li, Y.-L. Chen, Y. Li, and Q. Gong, Strongly enhanced light-matter interaction in a hybrid photonic-plasmonic resonator, Phys. Rev. A {\bf 85}, 031805(R) (2012).

\bibitem{Yin2016LB} Y. Yin, S. Li, S. B\"{o}ttner, F. Yuan, S. Giudicatti, E. S. G. Naz, L. Ma, and O. G. Schmidt, Localized Surface Plasmons Selectively Coupled to Resonant Light in Tubular Microcavities, Phys. Rev. Lett. {\bf 116}, 253904 (2016).

\bibitem{Conteduca2017RS} D. Conteduca, C. Reardon, M. G. Scullion, F. Dell’Olio, M. N. Armenise, T. F. Krauss, C. Ciminelli, Ultra-high $Q/V$ hybrid cavity for strong light-matter interaction, APL Photonics {\bf 2}, 086101 (2017)

\bibitem{Cognee2019DL}  K.G. Cogn\'{e}e, H.M. Doeleman, P. Lalanne, and A. F. Koenderink, Cooperative interactions between nano-antennas in a high-$Q$ cavity for unidirectional light sources, Light: Science \& Applications {\bf 8}, 115 (2019).

\bibitem{Zhang2022LW} H. Zhang, Y.-C. Liu, C. Wang, N. Zhang, and C. Lu, Hybrid photonic-plasmonic nano-cavity with ultra-high $Q/V$, Opt. Lett.  {\bf 45},4794 (2020). 

\bibitem{Xiong2022Xiao} X. Xiong and Y.-F. Xiao, Hybrid plasmonic-photonic microcavity for enhanced light-matter interaction, Sci. Bulletin {\bf 67}  1205 (2022).


\bibitem{Badolato20005HA}  A. Badolato, K. Hennessy, M. Atat\"{u}re, J. Dreiser, E. Hu, P. M. Petroff, and A. Imamo\v{g}lu, Deterministic Coupling of Single Quantum Dots to Single Nanocavity Modes, Science {\bf 308}, 1158 (2005).


\bibitem{Englund2007FF} D. Englund, A. Faraon, I. Fushman, N. Stoltz, P. Petroff, and J. Vu\v{c}kovi\'{c}, Controlling cavity reflectivity with a single quantum dot, Nature {\bf 450}, 857 (2007).

\bibitem{Bose2012SK} R. Bose, D. Sridharan, H. Kim, G.S. Solomon, and E. Waks, Low-Photon-Number Optical Switching with a Single Quantum Dot Coupled to a Photonic Crystal Cavity, Phys. Rev. Lett. {\bf 108}, 227402 (2012).

\bibitem{Choi2017HE} H. Choi, M. Heuck, and D. Englund, Self-Similar Nanocavity Design with Ultrasmall Mode Volume for Single-Photon Nonlinearities, Phys. Rev. Lett. {\bf 118}, 223605 (2017).

\bibitem{Najer2019SS} D. Najer, I. S\"{o}llner, P. Sekatski, et al., A gated quantum dot strongly coupled to an optical microcavity, Nature  {\bf 575}, 622–627 (2019). 

\bibitem{Albrechtsen2022LC} M. Albrechtsen, B. V. Lahijani, R. E. Christiansen, et al., Nanometer-scale photon confinement in topology-optimized dielectric cavities, Nat. Communications {\bf 13},  6281 (2022).


\bibitem{Gallagher2005} T.F. Gallagher, \emph{Rydberg Atoms} (Cambridge University Press, Cambridge, UK, 2005).

\bibitem{Saffman2010WM} M. Saffman, T. G. Walker, and K. M\o lmer, Quantum information with Rydberg atoms, Rev. Mod. Phys. {\bf 82}, 2313 (2010).

\bibitem{Guerlin2010BE} C. Guerlin, E. Brion, T. Esslinger, and K. M\o lmer, Cavity quantum electrodynamics with a Rydberg-blocked atomic ensemble, Phys. Rev. A {\bf 82}, 053832 (2010).

\bibitem{Parigi2012BS}  V. Parigi, E. Bimbard, J. Stanojevic, A. J. Hilliard, F. Nogrette, R. Tualle-Brouri, A. Ourjoumtsev, and P. Grangier, Observation and Measurement of Interaction-Induced Dispersive Optical Nonlinearities in an Ensemble of Cold Rydberg Atoms, Phys. Rev. Lett. {\bf 109}, 233602 (2012).

\bibitem{Zhang2013SW}  X.-F. Zhang, Q. Sun, Y.-C. Wen, W.-M. Liu, S. Eggert, and A.-C. Ji, Rydberg Polaritons in a Cavity: A Superradiant Solid, Phys. Rev. Lett. {\bf 110}, 090402 (2013).

\bibitem{Haroche2020BR} S. Haroche, M. Brune, and J.M. Raimond,  From cavity to circuit quantum electrodynamics, Nat. Phys. {\bf 16}, 243–246 (2020).


\bibitem{Svelto2010} O. Svelto, \emph{Principles of Lasers} (Springer, New York, 2010).

\bibitem{Yun2001Chang} T.-Y. Yun and K. Chang, Uniplanar one-dimensional photonic-bandgap structures and resonators,  IEEE Trans. Microw. Theory Tech, {\bf 49}, 549-553 (2001).

\bibitem{Armenise2010CC} M.N. Armenise, C.E. Campanella, C. Cimieri, F. Dell’Olio, and V.M.N. Passaro, Phononic and photonic band gap structures: modelling and applications, Physics Procedia {\bf 3}, 357 (2010).

\bibitem{Molesky2018LP} S. Molesky,  Z. Lin, A. Y. Piggott, et al. Inverse design in nanophotonics. Nat. Photonics {\bf 12}, 659–670 (2018).

\bibitem{Albrechtsen2022L} M. Albrechtsen, B.V. Lahijani, R.E. Christiansen, et al., Nanometer-scale photon confinement in topology-optimized dielectric cavities, Nat. Communications {\bf 13},  6281 (2022). 

\bibitem{supp} See the Supplemental Material for discussions of the properties of the plane-plane Fabry–P\'{e}rot cavity,  the winged symmetric confocal Fabry–P\'{e}rot cavities with two different sizes, and the winged microdome Fabry–P\'{e}rot cavities, as well as the analyses of the local coupling strength and Purcell enhancement using both the simplified single high-$Q$ mode approximation and the Green’s dyadic approach, which includes Ref.\,\cite{Lalanne2018YV}.

\bibitem{Lalanne2018YV} P. Lalanne, W. Yan, K. Vynck, C. Sauvan, J.-P. Hugonin, Light Interaction with Photonic and Plasmonic Resonances, Laser Photonics Rev.  {\bf 12}, 1700113 (2018).

\bibitem{Kristensen2012VVH} P.T. Kristensen, C.Van Vlack, and S. Hughes, Generalized effective mode volume for leaky optical cavities, Opt. Lett. {\bf 37},  1649-1651 (2012).

\bibitem{Kristensen2014Hughes} P.T. Kristensen and S. Hughes, Modes and Mode Volumes of Leaky Optical Cavities and Plasmonic NanoresonatorsArticle link copied, ACS Photonics  {\bf 1}, 2(2014).

\bibitem{Kristensen2020HI} P.T. Kristensen, K. Herrmann, F. Intravaia, and K. Busch, Modeling electromagnetic resonators using quasinormal modes, Adv. Opt. Photonics {\bf 12}, 612-708 (2020).


\bibitem{Spillane2005KV} S.M. Spillane, T.J. Kippenberg, K.J. Vahala, K.W. Goh, E. Wilcut, and H.J. Kimble, Ultrahigh-Q toroidal microresonators for cavity quantum electrodynamics, Phys. Rev. A {\bf 71}, 013817 (2005).

\bibitem{Srinivasan2006BP} K. Srinivasan, M. Borselli, O. Painter, A. Stintz, and S. Krishna, Cavity Q, mode volume, and lasing threshold in small diameter AlGaAs microdisks with embedded quantum dots, Opt. Express {\bf 14}, 1049 (2006).

\bibitem{Buck2003Kimble}  J.R. Buck and H.J. Kimble, Optimal sizes of dielectric microspheres for cavity QED with strong coupling, Phys. Rev. A {\bf 67}, 033806 (2003).

\bibitem{Vernooy1997Kimble}  D.W. Vernooy and H.J. Kimble, Quantum structure and dynamics for atom galleries, Phys. Rev. A {\bf 55} 1239 (1997). 

\bibitem{Stanley1994HO}  R. P. Stanley,  R. Houdr\'{e},  U. Oesterle,  M. Gailhanou, and  M. Ilegems, Ultrahigh finesse microcavity with distributed Bragg reflectors, Appl. Phys. Lett. {\bf 65}, 1883 (1994)

\bibitem{Xu2019WC} X.Q. Wu, Y.P. Wang, Q.S. Chen, Y.-C. Chen, X.Z. Li, L.M. Tong, and X.D. Fan, High-Q, low-mode-volume microsphere-integrated Fabry–Perot cavity for optofluidic lasing applications, Photon. Res. {\bf 7}, 50 (2019).

\bibitem{Trichet2016TD} A. A. P. Trichet, P. R. Dolan, D. James, G. M. Hughes, C. Vallance, and J. M. Smith, Nanoparticle Trapping and Characterization Using Open Microcavities, Nano Lett. {\bf 16}, 6172 (2016).

\bibitem{Gaponenkoa2019KR}  N.V. Gaponenkoa, P.A. Kholova, T.F. Raichenokb, and S.Ya Prislopskib, Enhanced luminescence of europium in sol-gel derived BaTiO3/SiO2 multilayer cavity structure, Opt. Materials {\bf 96}, 109265 (2019).

\bibitem{Shihab2020AM}  N.K. Shihab, J. N. Acharyya, U.P. Mohammed Rasi, R.B. Gangineni, G. Vijaya Prakash, and D. Narayana Rao, Cavity enhancement in nonlinear absorption and photoluminescence of BaTiO3, Optik  {\bf 207}, 163896 (2020). 


\bibitem{Gu2017KM} X. Gu, A. F. Kockum, A. Miranowicz, et al., Microwave photonics with superconducting quantum circuits, Phys. Rep. {\bf 718–719}, 1 (2017).

\bibitem{Kockum2019MD} A.F. Kockum, A. Miranowicz, S. De Liberato, et al., Ultrastrong coupling between light and matter, Nat. Rev. Phys. {\bf 1}, 19 (2019).

\bibitem{FornDiaz2019LR} P. Forn-D\'{i}az, L. Lamata, E. Rico, et al., Ultrastrong coupling regimes of light-matter interaction, Rev. Mod. Phys. {\bf 91}, 025005 (2019).

\bibitem{Birnbaum2005BM} K.M. Birnbaum, A. Boca, R. Miller, A.D. Boozer, T.E. Northup, and H.J. Kimble, Nature {\bf 436}, 87 (2005).


\bibitem{Miranowicz2013PL} A. Miranowicz, M. Paprzycka, Y.-x. Liu, J. Bajer, and F. Nori, Two-photon and three-photon blockades in driven nonlinear systems, Phys. Rev. A {\bf 87}, 023809 (2013).

\bibitem{Hamsen2017TW} C. Hamsen, K. N. Tolazzi, T. Wilk, and G. Rempe,  Two-Photon Blockade in an Atom-Driven Cavity QED System, Phys. Rev. Lett. {\bf 118}, 133604 (2017).

\bibitem{Lounis2005Orrit} B. Lounis and M. Orrit, Single-photon sources, Rep. Prog. Phys. {\bf 68} 1129 (2005). 

\bibitem{Glauber1963} R. J. Glauber, The quantum theory of optical coherence, Phys. Rev. {\bf 130}, 2529 (1963).

\bibitem{Scully1997Zubairy} M.O. Scully and M.S. Zubairy, {\it Quantum Optics} (Cambridge University Press, Cambridge, England, 1997).

\bibitem{Agarwal2013} G.S. Agarwal, {\it Quantum optics} (Cambridge University Press, 2013).

\bibitem{Johansson2012NN} J.R. Johansson, P.D. Nation, and F. Nori, QuTiP: An open- source Python framework for the dynamics of open quantum systems, Comput. Phys. Commun. {\bf 183}, 1760 (2012).

\bibitem{Ladd2010JL} T. D. Ladd, F. Jelezko, R. Laflamme, Y. Nakamura, C. Monroe, and J. L. O’Brien, Quantum computers, Nature  {\bf 464}, 45 (2010). 

\bibitem{Gisin2002Thew}  N. Gisin and R. Thew,  Quantum communication, Nat. Photonics {\bf 1}, 165 (2007).

\bibitem{Bassoli2021BD} R. Bassoli, H. Boche, C. Deppe, R. Ferrara, F. H. P. Fitzek, G. Janssen, and S. Saeedinaeeni, \emph{Quantum Communication Networks}, Springer (2021).

\bibitem{Kolobov2007} M.I. Kolobov, \emph{Quantum Imaging},  Springer (2007).

\bibitem{Magana-Loaiza2019Boyd} O.S. Maga\~{n}a-Loaiza and R.W. Boyd, Quantum imaging and information, Rep. Prog. Phys. {\bf 82}, 124401(2019).















































































 
\end{thebibliography}
\end{document}